\documentclass[
    ,final            
  ]
  {aipproc}

\layoutstyle{6x9}
\usepackage{amssymb}

\begin{document}

\title{Multiwavelength Gamma-Ray Bursts Observations with ECLAIRs}


\author{Diego G\"{o}tz on behalf of the ECLAIRs collaboration (CEA Saclay, CESR Toulouse, APC Paris, MIT Boston)}{
  address={CEA -- SAp/DAPNIA/DSM, Orme des Merisiers, F-91191, Gif sur Yvette, France}
,altaddress={e-mail: diego.gotz@cea.fr}
}



\begin{abstract}
ECLAIRs is the next space borne instrument that will be fully dedicated
to multi-wavelength studies of Gamma-Ray Bursts (GRBs).
It consists of a coded mask telescope with a wide ($\sim$2 sr) field of view, made
of 6400 CdTe pixels ($\sim$1000 cm$^{2}$), which will
work in the 4--300 keV energy band.
It is expected to localise $\sim$80 GRBs/yr, thanks to the on-board real
time event processing. The GRBs (and other transients) coordinates
will be distributed within a few seconds from the onset
of the burst with a typical uncertainty of $\sim$5--10 arcmin.
The detection system will also include a soft X-ray
camera (1--10 keV) allowing to study in detail the prompt soft X-ray emission
and to reduce the error box for about half of the
GRBs seen by ECLAIRs to $\sim$30 arcsec.
ECLAIRs is expected to be flown in late 2011 and to be the only instrument capable
of providing GRB triggers with sufficient localisation accuracy for
GRB follow-up observations with the powerful ground based
spectroscopic telescopes available by then.
We will present the current status of the ECLAIRs project and its possible
developments.
\end{abstract}

\maketitle


\section{Introduction}
Ever since their discovery \cite{klebesadel},
Gamma-Ray Bursts (GRBs) have been a puzzling mystery, mostly because
of their short durations and the apparent
lack of counterparts at other wavelengths. A breakthrough
in this field came thanks to the Italian-Dutch
satellite {\it Beppo}SAX, which had the capability to
localise the bursts' prompt emission  with a
precision of a few arcminutes within a few hours. This led to the discovery
of the afterglow emission at lower energies, initially in X-rays
\cite{costa} and subsequently at optical \cite{vanpa}
and radio \cite{frail} wavelengths, which allowed 
the redshift of these objects to be measured, and firmly established
the cosmological nature of GRBs.

Due to the limited duration and the fading
character of the afterglow emission, the prompt distribution
of GRBs coordinates to the scientific community is a high priority.
After the end of the {\it Beppo}SAX mission this task has been accomplished
mainly by {\it HETE-II} \cite{ricker}, and {\it INTEGRAL} through
to the INTEGRAL Burst Alert System (IBAS; \cite{ibas}).
Since 2005, thanks to the {\it Swift} satellite \cite{swift}, and to its capability to pinpoint the postions of GRBs to a few arcseconds in few minutes, the GRB field is progressing quickly. In particular the association between long GRBs and the death of massive ($\gtrsim$20 M$_{\odot}$) stars is becoming stronger, and the study of the early X-ray afterglows has shown new flaring phases, which were unknown before {\it Swift}. In addition, thanks to its good sensitivity, {\it Swift} is localising a growing number of GRBs at high redshift, the most distant one being 050904 at a redshift of 6.29 \cite{watson}.

Although considerable progress in the understanding of GRBs has been made in the last years, some crucial points still need some further investigation: the nature of the prompt emission, its relationship to the afterglow, the nature of the central engine, the nature of the progenitor star, in particular in connection to cosmological studies. In order to answer to these questions, it is fundmental to collect as much information as possible on every single GRB. The ECLAIRs project is being developed for this scope.

\section{The ECLAIRs project}
The goal of the ECLAIRs project is conceived to detect during its lifetime at least 200 GRBs, of all types, short and long, and in particular the ones with a soft spectrum, potentially the most distant ones \cite{schanne05,schanne06}. The GRBs should be observed also before and after the main event, in particular to study the spectrum of a possible precursor. These GRBs should be observed on a broad energy range in order to study the prompt X-ray component, and to constrain as much as possible the bursts ``peak energy", $E_{p}$, namely the energy at which a GRB emits the bulk of its energy.
The determination of $E_{p}$ is very important since in the last years some correlations have been found between this quantity and the isotropic equivalent energy, $E_{iso}$ \cite{amati}, and between $E_{p}$ and $E_{iso}(1-cos\;\theta)$, the collimation corrected  energy \cite{ggl04}. The scatter around the latter correlation is tight, and suggests to use the GRBs as cosmological tools. Indeed Ghirlanda \& Ghisellini [2004] estimate that a sample of 12 well studied GRBs with redshifts between 0.9 and 1.1 are enough to calibrate their correlation, and its application to 150 bursts gives good constraints on the cosmological parameters $\Omega_{M}$ and $\Omega_{\Lambda}$. These constraints are complementary in the ($\Omega_{M}$, $\Omega_{\Lambda}$) plane to the ones obtained from the cosmic microwave background and from the Supernovae of type I.

 Another fundamental aspect of the project is to ensure the possibility of a follow-up from large ground based optical telescopes for at least 75\% of the detected bursts. The observation plan will in fact be adjusted in order to have at least one 8-m-class telecope that covers a large fraction of ECLAIRs field of view. This is crucial in order to be able to determine the redshift of GRBs, and hence their luminosity, in particular, in view of the second generation spectrometers that will be available in 2011, like X-Shooter.
 In order to allow for the follow up, ECLAIRs will have to provide a coarse (10 arcmin) localisation within 10 s from the burst onset. This information can be exploited from ground based robotic telescopes. For half of the GRBs a position with an accuracy of 30 arcsec should be provided within the same time interval. Within 1 minute the best available information should reach all the major telescopes on Earth.


\subsection{The Instruments}
\begin{figure}
  \includegraphics[height=0.5\textheight]{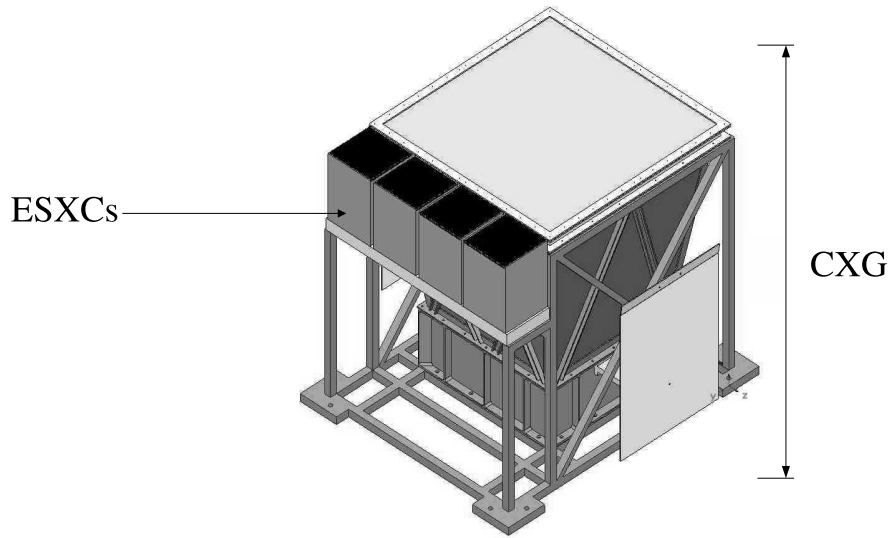}
  \caption{The ECLAIRs telescope}
  \label{fig:ecl}
\end{figure}

The ECLAIRs telescope will be composed by the X and Gamma-Ray Camera (CXG, the trigger device) and a Soft X-ray Camera (ESXC), see Fig. \ref{fig:ecl}. Besides the space borne instruments the mission includes also two dedicated Ground Follow-up robotic Telescopes (GFT) to study the prompt and early afterglow emission of GRBs.

The CXG and ESXC will have a wide field of view of $\sim$2 sr. They are both coded mask telescopes and will be operating in th 4--300 keV and 1--10 keV energy range respectively. The CXG will be made of 80$\times$80 pixels of CdTe yielding an effective area of 1024 cm$^{2}$, and will be able to localise all the bursts in its field of view within 10 arcmin (for a 5$\sigma$ source). The ESXC camera, triggered by the CXG with its coarse position, will be able to localise the prompt GRBs emission, for about half of the burst, within 30 arcsec (5 $\sigma$).  This instrument will be based on the experience of the soft X-ray camera on board {\it HETE-II}.
The characteristics of the ECLAIRs instruments are summarised in Table \ref{tab:hien}.

\begin{table}
\caption{Characteristics of ECLAIRs instruments}
\begin{tabular}{|c|c|c|}
\hline  & ECLAIRs/CXG & ECLAIRs/ESXC  \\ 
\hline Energy range & 4--300 keV & 1--10 keV  \\ 
\hline Field of view & 2 sr & 2 sr  \\ 
\hline Sensitive area & 1024 cm$^{2}$ CdTe  & 96 cm$^{2}$ Si\\ 
\hline Mask open fraction & 30\% & 20\%  \\ 
\hline Temporal resolution & 10 $\mu$s & 100 ms \\ 
\hline Source localisation & 10 arcmin for 5$\sigma$ & 30 arcsec for 5$\sigma$  \\ 
\hline 
\end{tabular} 
\label{tab:hien}
\end{table}

In the framework of the ECLAIRs project the addition of more instruments is being studied in order to extend the energy band over which the GRBs can be studied: as stated earlier, the measurement of $E_{p}$ is of primary importance. Hence a gamma-ray detector, extending the energy range up to a few MeV, is currently being studied. Besides the GFT a wide field optical camera could be placed on board in order to study the prompt optical emission.


\subsection{Number of GRBs per year}
With the current mission design one expects that ECLAIRs will localise about 80 GRBs per year. This calculation is based on the instrument geometry, and on a background evaluation made under the hypothesis that between 4 and 50 keV (the CXG triggering band) the background is dominated, in absence of strong sources, by the cosmic diffuse X-ray background.
An analytic expression of it can be found in \cite{gruber}, and is
\begin{equation}
7.877\times E^{-0.29} e^{-E/41.13}\; \rm{\frac{keV}{keV\, cm^{2}\, s\, sr}}
\label{extra}
\end{equation}
By numerically integrating Eq. \ref{extra} between 4 and 50 keV, multiplying for the detector geometrical area (A$_{eff}$=0.4$\times$0.4$\times$80$\times$80 cm$^{2}$ = 1024 cm$^{2}$), and taking the aperture and the average efficiency into account (the CdTe efficiency is practically 100\% in this energy range), one obtains a background value of 2050 counts s$^{-1}$ sr$^{-1}$.
To evaluate the number of GRBs in the sky we numerically integrated the peak flux density distribution derived over 1 s with BATSE \cite{4th}, N(P), where N is the number of bursts detected per year and above a given peak flux P.
 We then took into account the CXG livetime which is mainly limited by the passages above the southern Atlantic anomaly and by the time the earth enters the field of view (see below). In the hypothesis of a pure antisolar pointing and an equatorial orbit with an inclination of 30$^{\circ}$ one obtains, on average, that for 64\% of the orbit we have a field of view not occulted by the earth. We will call this factor, $lt$=0.59. The triggering energy band of the CXG is different from the BATSE one (50-300 keV). So, first of all one has to convert the BATSE fluxes into CXG fluxes and this has been done by assuming a GRB spectrum, described by a Band function \cite{band}, with average parameters ($E_{p}$=200 keV, $\beta$=--2.25, $\alpha$=--1) \cite{preece}. This introduces a multiplicative factor of $ec$=2.45 on the Log N-Log P. In addition recent missions like {\it Beppo}SAX, {\it HETE-II}, {\it INTEGRAL} and {\it Swift}, have shown that there is an entire class of GRBs, the so called, X-Ray Flashes (XRFs), that were too soft to be detected by BATSE. So we have to add another multiplicative factor that takes this into account, and its value is $n_{X}\sim$1.55 (=(N$_{GRB}$+N$_{XRR}$+N$_{XRF}$)/(N$_{GRB}$+N$_{XRR}$)), as derived by {\it HETE-II} data \cite{sakamoto}.
 
 The CXG sensitivity limit in the 4-50 keV band for a 5 $\sigma$ detection in the totally coded field of view is $S_{lim0}\sim$0.7 ph cm$^{-2}$ s$^{-1}$, based on the background value derived above and taking into account that a square cm of detector sees a different sky fraction depending on its position on the detector (a variation between $\sim$0.9 and $\sim$ 1 sr).
 
 By taking the above factors, the variation of the effective area as a function of the off-axis angles of the instrument A$_{eff}$($\theta_{X}$,$\theta_{Y}$), the relative sensitivity variation, S$_{lim}$($\theta_{X}$,$\theta_{Y}$, and the variation of the solid angle seen by the detector in each direction of the field of view, $\Omega$($\theta_{X}$,$\theta_{Y}$), into account, one derives the number of GRBs per year observable with the CXG in the following way:

\begin{equation}
N_{GRB}=lt\; n_{X} \int_{\theta_{X}} \int_{\theta_{Y}}S_{lim}(\theta_{X},\theta_{Y}) \cdot \Omega(\theta_{X},\theta_{Y})/4\pi\, \int_{S_{lim}(\theta_{X},\theta_{Y})}^{\infty} N(P\cdot ec) dP\; d\theta_{X}\, d\theta_{Y}\, 
\label{ngrb}
\end{equation}

with $\theta_{X}$ and $\theta_{Y}$ being the off-axis angles that describe the field of view. The result is 76 bursts/year.

\subsection{The Alert Strategy}

The positions of the transients detected by the ECLAIRs telescope will be derived on board. A dedicated real-time processing unit, UTS, will be present. The UTS will look for significant excesses in the CXG light curve simultaneously on different time scales, energy bands and detector portions. In case of a positive excess the imaging process will be invoked and, if a previously uncatalogued source is found, an alert will be generated. This alert will be transmitted promptly to the ground, so that robotic telescopes can quickly react. At the same time the infomation will be passed to the ESXC in order to refine the position, if possible. At the same time images on different energy bands and timescales are produced continuously and compared to previously recorded background images in order to spot new transients that could have been missed by the previous algorithms. The latter,
introduced for the first time in the IBAS system, is particularly effective in detecting long and ``slowly rising" bursts, like 050904.
The positions derived on board with ECLAIRs will be distributed to the ground via a VHF network, which will be an extension of the one used today by {\it HETE-II}. 

\subsection{The Follow-up}
  The pointing strategy and the orbit of ECLAIRs will be adapted in order to have a large chance of follow-up by the largest ground based telescopes. The launcher should place ECLAIRs into a circular orbit at 600 km altitude with an inclination of 30$^{\circ}$. During most of the orbit the satellite will perform an anti-solar pointing. 
In addition the avoidance of regions of bright X-ray sources, in particular Sco X-1, can be accommodated by varying the angle between the orbital plane and the pointing direction.

\section{Conclusions}
The ECLAIRs system could be incorporated in a next mission dedicated to GRBs. Besides GRBs, this mission will include the study of all the transient phenomena in the high-energy sky, such as X-ray bursts, Soft Gamma-Ray Repeaters, AGNs, Supernovae, Novae, etc. 

Although thanks to {\it Swift} considerable progress has been made in GRB science, some aspects will be explored further thanks to ECLAIRs. These aspects include \begin{itemize}
\item the study of the prompt X-ray emission associated to GRBs
\item the detection of a large sample of soft GRBs, potentially the most distant ones, thanks to the 4--50 keV triggering band 
\item the broad band spectra (from 1 keV to a few MeV)
\item the prompt (and/or precursive) optical emission
\end{itemize}
In addition the ECLAIRs pointing strategy will be crucial to ensure the most favourable conditions for follow-up studies from ground based telescopes.

\begin{theacknowledgments}
  DG acknowledges support from the French Space Agency (CNES).
\end{theacknowledgments}



\bibliographystyle{aipproc}   


\begin{thebibliography}{7}
\bibitem[Amati et al. 2002]{amati}Amati, L., et al. 2002, A\&A, 390, 81
\bibitem[Band et al. 1993]{band} Band D., et al. 1993, ApJ, 413, 281
\bibitem[Barthelmy et al. 2005]{swift}Barthelmy, S. D., et al. 2005, Space Science Reviews, 120, 143 
\bibitem[Costa et al. 1997] {costa} Costa, E., Frontera, F., Heise, J., et al. 1997, Nature, 387, 783
\bibitem[Frail et al. 1997]{frail} Frail, D. et al. 1997, Nature, 389, 261
\bibitem[Ghirlanda et al. 2004]{ggl04}Ghirlanda, G., Ghisellini, G., \& Lazzati, D. 2004, ApJ, 616, 331
\bibitem[Ghirlanda \& Ghisellini 2004]{gg04}Ghirlanda, G., \& Ghisellini, G. 2004, in proceedings of the workshop "Science with the New Generation of High-Energy Gamma-Ray Experiments", Cividale del Friuli (Italy), astro-ph/0602498
\bibitem[Gruber et al. 1999]{gruber} Gruber D.E., et al. 1999, ApJ, 520, 124
\bibitem[Klebesadel et al. 1973]{klebesadel} Klebesadel, R. et al. 1973, ApJ, 182, L85
\bibitem[Mereghetti et al. 2003]{ibas} Mereghetti, S., G\"otz, D., Borkowski, J., et al. 2003, A\&A, 411, L291 
\bibitem[Paciesas et al. 1999]{4th} Paciesas W.S., et al. 1999  ApJS, 122, 465
\bibitem[Preece 2000]{preece} Preece R.D. et al., 2000, ApJS, 126, 19
\bibitem[Ricker et al. 2002]{ricker} Ricker, G., Hurley, K., Lamb, D., et al. 2002, ApJ, 571, L127
\bibitem[Sakamoto et al. 2005]{sakamoto} Sakamoto T. et al., 2005, ApJS, 629, 311
\bibitem[Schanne et al. 2005]{schanne05}Schanne, S., Atteia, J.-L., Barret, D., et al. 2005, IEEE Transact. Nucl. Sci, 52, 2778
\bibitem[Schanne et al. 2006]{schanne06}Schanne, S., Atteia, J.-L., Barret, D., et al. 2006, NIM-A, in press
\bibitem[van Paradijs et al. 1997]{vanpa} van Paradijs, J., Groot, P.J., Galama, T., et al. 1997, Nature, 386, 686
\bibitem[Watson et al. 2006]{watson}Watson, D., Reeves, J.N., Hjorth, J, et al. 2006, ApJ, 637, L69
\end{thebibliography}


\end{document}